\documentclass[sigconf]{acmart}
\AtBeginDocument{%
  }

\setcopyright{acmlicensed}
\copyrightyear{2026}
\acmYear{2026}
\acmDOI{XXXXXXX.XXXXXXX}
\acmConference[MM'26]{Make sure to enter the correct
  conference title from your rights confirmation email}{November 10–14,
  2026}{Rio de Janeiro, Brazil}
\acmISBN{978-1-4503-XXXX-X/2025/06}
\usepackage{booktabs}   
\usepackage{tabularx}   
\usepackage{amsmath}    
\usepackage{array}       
\usepackage{ragged2e}    
\usepackage{graphicx} 
\usepackage{makecell}
\usepackage{multirow}
\usepackage{float} 
\usepackage{caption}
\usepackage{subcaption} 

\begin{document}

\title{MATHDance: Mamba-Transformer Architecture with Uniform \\ Tokenization for High-Quality 3D Dance Generation}

\author{Kaixing Yang}
\authornote{Equal Contribution.}
\email{yangkaixing@ruc.edu.cn}
\affiliation{%
    \institution{
        Renmin University of China
    }
    \city{Beijing}
    \country{China}
}

\author{Xulong Tang}
\authornotemark[1]
\email{xulong.tang@maloutech.com}
\affiliation{%
    \institution{
        Malou Tech Inc \\
    \state{Texas}
    \country{USA}    
    }
}

\author{Ziqiao Peng}
\authornotemark[1]
\email{pengziqiao@ruc.edu.cn}
\affiliation{%
    \institution{
        Renmin University of China \\
    \city{Beijing}
    \country{China}  
    }
}

\author{Yuxuan Hu}
\email{huyuxuan1999@ruc.edu.cn}
\affiliation{%
    \institution{
        Renmin University of China
    }
    \city{Beijing}
    \country{China}
}


\author{Xiangyue Zhang}
\email{xiangyuezhang@whu.edu.cn}
\affiliation{%
    \institution{
        Wuhan University
    }
    \city{Wuhan}
    \country{China}
}

\author{Puwei Wang}
\authornote{Corresponding author.}
\email{wangpuwei@ruc.edu.cn}
\affiliation{%
    \institution{
        Renmin University of China
    }
    \city{Beijing}
    \country{China}
}

\author{Jun He}
\authornotemark[2]
\email{hejun@ruc.edu.cn}
\affiliation{%
    \institution{
        Renmin University of China
    }
    \city{Beijing}
    \country{China}
}

\author{Hongyan Liu}
\authornotemark[2]
\email{liuhy@sem.tsinghua.edu.cn}
\affiliation{%
    \institution{
        Tsinghua University
    }
    \city{Beijing}
    \country{China}
}

\author{Zhaoxin Fan}
\authornotemark[2]
\email{zhaoxinf@buaa.edu.cn}
\affiliation{%
    \institution{
        Beihang University
    }
    \city{Beijing}
    \country{China}
}

\renewcommand{\shortauthors}{Anonymous et al.}

\begin{abstract}

Music-to-dance generation represents a challenging yet pivotal task at the intersection of choreography, virtual reality, and creative content generation. Despite its significance, existing methods face substantial limitation in achieving choreographic consistency. To address the challenge, we propose MatchDance, a novel framework for music-to-dance generation that constructs a latent representation to enhance choreographic consistency. MatchDance employs a two-stage design: (1) a Kinematic-Dynamic-based Quantization Stage (KDQS), which encodes dance motions into a latent representation by Finite Scalar Quantization (FSQ) with kinematic-dynamic constraints and reconstructs them with high fidelity, and (2) a Hybrid Music-to-Dance Generation Stage(HMDGS), which uses a Mamba-Transformer hybrid architecture to map music into the latent representation, followed by the KDQS decoder to generate 3D dance motions. Additionally, a music-dance retrieval framework and comprehensive metrics are introduced for evaluation. Extensive experiments on the FineDance dataset demonstrate state-of-the-art performance. Code will be released upon acceptance.

\end{abstract}



\received{20 February 2007}
\received[revised]{12 March 2009}
\received[accepted]{5 June 2009}


\begin{teaserfigure}
  \centering
  \includegraphics[width=\linewidth]{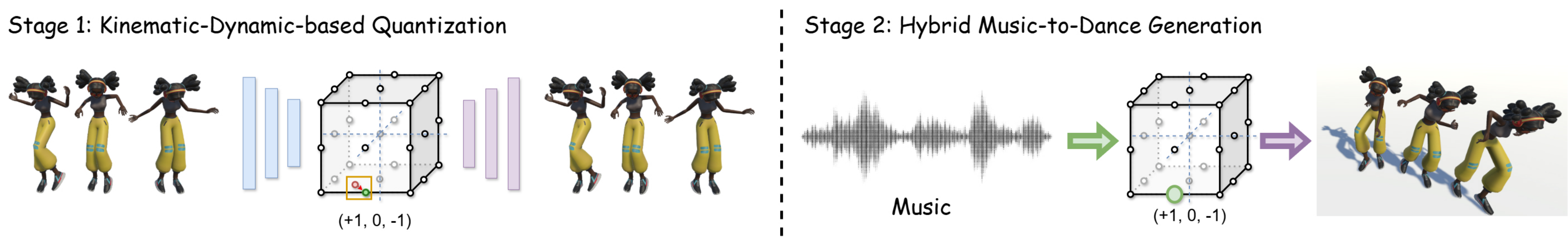}
  \caption{To enhance choreographic consistency, MATHDance designs High-Fidelity Dance Tokenization stage for physical plausibility and Hybrid Music-to-Dance Generation stage for aesthetic quality.}
  \label{fig:teaser}
  \vspace{0.2in}
\end{teaserfigure}

\maketitle

\section{Introduction}

Music-to-dance generation is a crucial task that translates auditory input into dynamic motion, with significant applications in virtual reality, choreography, and digital entertainment \cite{li2023finedance,li2021ai,yang2025mace}. By automating this process, it enables deeper exploration of the intrinsic relationship between audio and movement~\cite{zhang2025robust,zhang2025semtalk,siyao2023bailando++}, while expanding possibilities for creative content generation\cite{zhou2026framesequalcomplexityawaremasked,zhang2025echomask}. 

Current music-to-dance generation approaches have witnessed rapid progress and can be broadly categorized into two paradigms\cite{siyao2022bailando,zhuang2023gtn}: (1) One-stage methods directly map musical features to human motion\cite{li2024lodge,li2024lodge++,huang2020dance}. (2) Two-stage methods first construct choreographic units and then learn their probability distributions conditioned on music~\cite{siyao2022bailando,zhang2026mitigating,zhang2025,siyao2024duolando,yang2026tokendancetokentotokenmusictodancegeneration}. However, previous methods suffer from issues with choreographic consistency, resulting in a lack of physical plausibility and poor aesthetic quality. For example, the generated motions often contain unnatural limb movements, and exhibit repetitive or overly stationary patterns. Simultaneously, current evaluation metrics fail to capture the artistic essence of music and dance, thereby misleading the generative process toward producing dances that lack choreographic consistency. For example, the mainstream FID, computed on handcrafted stylistic, kinetic, and geometric features~\cite{li2021ai,li2023finedance}, primarily reflects surface-level motion statistics and lacks sensitivity to deeper choreographic semantics. BAS~\cite{li2021ai,li2023finedance}, meanwhile, reduces music-dance correspondence to beat-level alignment, overlooking hierarchical rhythmic structures.

The core idea of MATHDance is to enhance choreographic consistency by decoupling it into two complementary aspects: physical plausibility and aesthetic quality. Physical plausibility refers to movements that follow natural body mechanics and remain smooth and stable over time. Aesthetic quality, on the other hand, focuses on how well the dance matches the music in rhythm, style, and structure. To address these, MATHDance adopts a two-stage architecture, as shown in Fig. \ref{fig:teaser}. (1) High-Fidelity Dance Tokenization (HFDT) focuses on physical plausibility. We introduce Finite Scalar Quantization (FSQ) for dance tokenization. Unlike the learned vector quantization in VQ-VAE, FSQ applies fixed uniform scalar quantization, leading to stable token usage and improved latent expressiveness. To further promote spatio-temporal coherence, we impose spatial constraints via Forward Kinematics (FK) reconstruction, and temporal constraints by considering velocity and acceleration during reconstruction. Together, HFDT effectively constrains the latent motion space to physically plausible regions. (2) Hybrid Music-to-Dance Generation (HMDG) aims to enhance aesthetic quality. We propose a Mamba-Transformer hybrid architecture that combines Mamba's efficient modeling of local dependencies with the Transformer's capacity for global context. Moreover, we utilize a Sliding Window Attention mechanism instead of traditional Casual Attention mechanism, to better align long-term dance generation scene. Furthermore, we leverage the Music Foundation Model MuQ\cite{zhu2025muq} learned from multi-level music informatics tasks, for powerful music representation. HMDG promotes choreography that aligns with music in rhythm, structure, and expression, supporting aesthetic quality.

On the other hand, we introduce a retrieval-based evaluation protocol, which leverages the intrinsic connection between retrieval and generation tasks and utilizes contrastive learning to encode cross-modal music-dance semantics. This protocol has shown strong effectiveness in similar domains such as text-to-motion\cite{guo2022generating}, video-to-music\cite{zhuo2023video}, and text-to-video generation\cite{liu2024evalcrafter}. By training a retrieval model on real-world data, it effectively captures artistic essence of music and dance. Specifically, the model uses temporal processing and downsampling layers in the encoders, with CLIP loss \cite{radford2021learning} applied to align cross-modal features. Subsequently, the retrieval-based feature complements standard metrics like FID and DIV\cite{li2021ai}, providing a deeper assessment of generative models.

In the music-to-dance generation, the contributions of this work are summarized as follows: (1) We present MATHDance, a two-stage framework that enhances choreographic consistency by decoupling it into physical plausibility and aesthetic quality. Extensive experiments on AIST++ and FineDance demonstrate its superiority in both generation quality and computational efficiency. (2) We propose a dance tokenization method that introduces FSQ with spatial-temporal reconstruction constraints. Additionally, we design a dance token generation method that utilizes a Mamba-Transformer backbone and leverage the Music Foundation Model MuQ for powerful music representations. (3) We introduce a retrieval-based evaluation protocol, and conduct robust experiments to validate its reliability.

\section{Related Work}
\subsection{One-Stage Music-to-Dance Generation}
Music and dance are inherently connected, motivating research on music-driven 3D dance generation, where musical features are used to predict human motion. Early works adopt encoder-decoder frameworks to generate entire motion sequences~\cite{lee2018listen,tang2018dance,le2023music,correia2024music,li2021ai}. In AIGC, Generative Adversarial Networks (GANs) have been applied to enhance realism in music-to-dance generation~\cite{chen2021choreomaster,huang2020dance,yang2024cohedancers}. More recently, Diffusion Models achieved notable performance in this domain~\cite{tseng2023edge,li2023finedance,li2024lodge,li2024lodge++,le2023controllable,yang2025flowerdance}, although their high sampling cost limits long-sequence generation efficiency. However, the above methods lack explicit spatial constraints, often leading to nonstandard poses that extend beyond the dancing subspace.

\subsection{Two-Stage Music-to-Dance Generation}
Two-stage approaches leverage the periodicity of dance by (1) quantizing motion into dance token, and (2) learning music-conditioned distributions over them. Since these token originate from real motion, such methods inherently favor physical plausibility. \textbf{(1) Dance Tokenization Stage.} Early works~\cite{ye2020choreonet,huang2022genre,chen2021choreomaster} rely on uniform segmentation, which is computationally inefficient. Later, VQ-VAE~\cite{gong2023tm2d,guo2022generating} enables learnable unit construction with reduced cost. Bailando~\cite{siyao2022bailando,siyao2023bailando++} further decouples upper/lower body units to expand unit capacity. Recent works~\cite{siyao2023bailando++,li2024lodge++,yang2025megadance} use detailed SMPL parameter instead of traditional 3D keypoints. However, they treat all joints equally, ignoring kinematic hierarchy of human body. \textbf{(2) Dance Generation Stage.} Choreomaster~\cite{chen2021choreomaster} and DanceRevolution~\cite{huang2020dance} adopt RNN variants, while later methods~\cite{siyao2022bailando,siyao2023bailando++} use cross-modal Transformers for improved temporal modeling and music-motion alignment. 

In conclusion, existing methods lack choreographic consistency, primarily due to two limitations: (1) VQ-VAE-based tokenization suffers from low codebook utilization, weakening physical plausibility; (2) Transformer-only architectures rely on positional encodings, offering weak inductive biases for the continuous nature of music and dance, thereby undermining aesthetic quality. To enhance choreographic consistency, MATHDance designs High-Fidelity Dance Tokenization stage for physical plausibility and Hybrid Music-to-Dance Generation stage for aesthetic quality.

\subsection{Evaluation for Music-to-Dance Generation}
Designing objective quantitative metrics for music-to-dance generation remains challenging due to its high subjectivity (aesthetic discrepancy). Early approaches \cite{li2022danceformer,tang2018dance} measure generation quality by computing MSE or MAE distances between generated and real dances. Subsequent works employ feature extractors to compute feature-level distances, using metrics such as FID for realism and DIV for motion diversity. Some methods~\cite{chen2021choreomaster} adopt self-reconstructing motion autoencoders as feature extractors, while others~\cite{lee2019dancing,li2023finedance} use genre classifiers trained on labeled dance data. Several approaches~\cite{li2021ai,siyao2022bailando,tseng2023edge} extract handcrafted kinetic~\cite{onuma2008fmdistance} and geometric~\cite{muller2005efficient} features to evaluate motion quality. Simultaneously, accurate modeling of music-dance correspondence is also crucial in this task. Existing BAS-based methods~\cite{li2021ai,huang2020dance} focus primarily on beat-level alignment, while overlooking the richer interplay between music and dance, including rhythmic structures and stylistic semantics.

In conclusion, current evaluation metrics fails to capture the artistic essence of music and dance, thereby misleading the generative process toward producing dances that lack choreographic consistency. To address this, we introduce a retrieval-based evaluation protocol, which leverages the intrinsic connection between retrieval and generation tasks and utilizes contrastive learning to encode cross-modal music-dance semantics.

\section{Methodology}
\subsection{Problem Definition}
Given a music sequence $M=\{m_0, m_1, ..., m_T\}$ and a dance genre label $g$, the goal is to generate a corresponding dance sequence $D=\{d_0, d_1, ..., d_T\}$. Each music feature $m_t$ is a 1024-dimensional MuQ representation~\cite{zhu2025muq}, while each dance feature $d_t = [\tau; \theta]$ consists of SMPL root translation $\tau$ and 6D joint rotation~\cite{zhou2019continuity,loper2023smpl}. We synchronize $M$ and $D$ at 30 FPS to ensure precise temporal alignment.

\begin{figure*}[t]
  \centering
  \includegraphics[width=\linewidth]{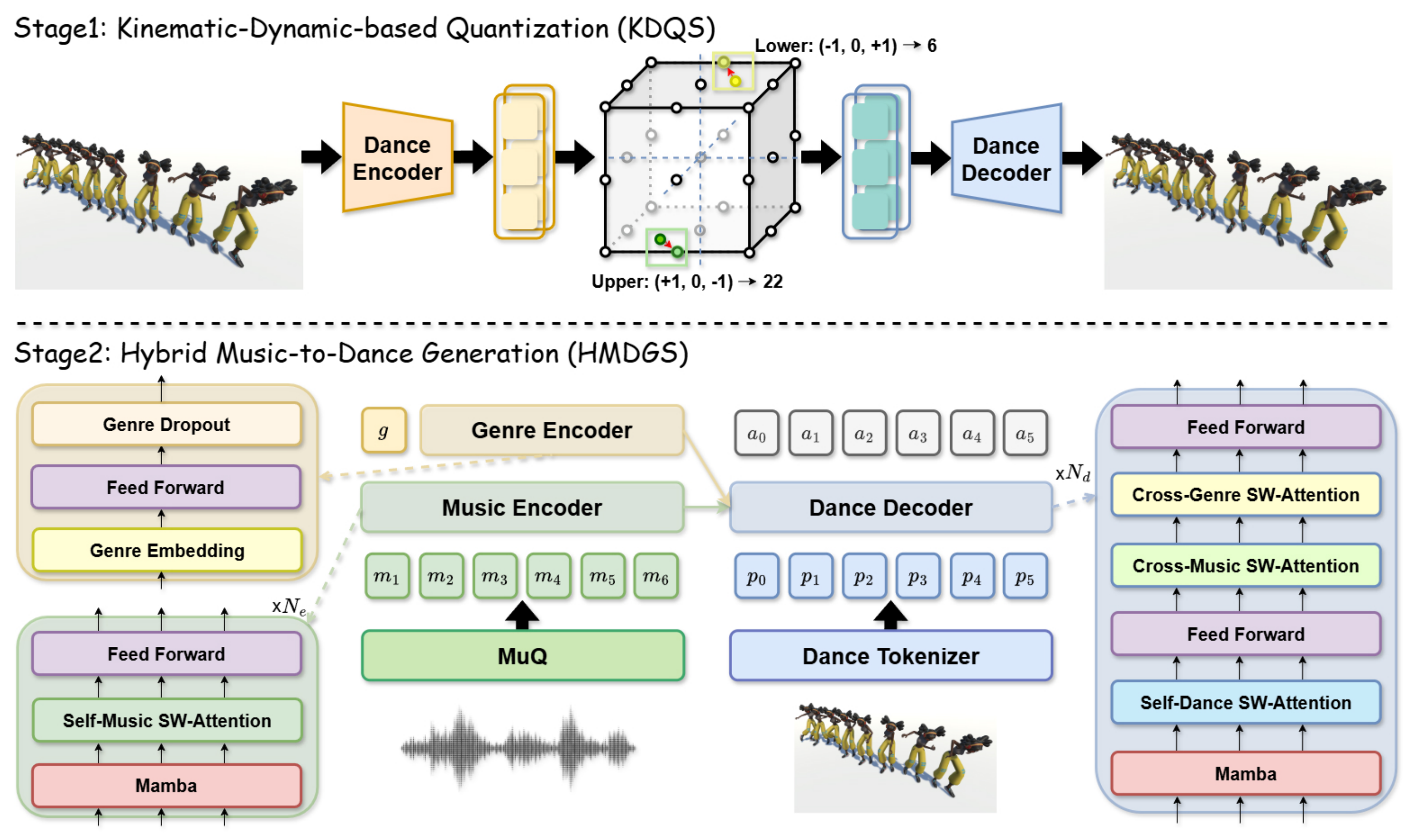}
  \caption{Overview of MATHDance. Stage 1 (HFDT) quantizes dance into upper/lower-body tokens by FSQ with spatial and temporal reconstruction constraints. Stage 2 (HMDG) autoregressively generates dance tokens from MuQ-derived music features. HMDG utilizes a Mamba-Transformer hybrid architecture, where Transformer is equipped with Sliding Window Attention (SWA). Residual connections are omitted for clarity.
  }
  \label{fig:overview}
\end{figure*}

\subsection{High-Fidelity Dance Tokenization}
Mainstream methods\cite{siyao2022bailando,siyao2023bailando++,siyao2024duolando} typically adopt VQ-VAE for motion tokenization. However, VQ-VAE often suffers from codebook collapse, where only a small subset of codes are utilized, leading to limited representational diversity and degrading physical plausibility of reconstructed dance \cite{mentzer2023finite}. To address this, we adopt Finite Scalar Quantization (FSQ), which replaces vector-wise code selection with differentiable scalar rounding. Unlike VQ-based methods that rely on discrete vector assignments, FSQ quantizes each feature dimension independently. This uniform tokenization scheme ensures not only balanced usage of code space, but also stable gradient flow during training.
 \subsubsection{Uniform Dance Tokenization.} Choreographic units form the basic elements of dance structure, exhibiting commonality across styles and tempos. We aim to learn a reusable codebook that unsupervisedly encodes any dance sequence into composable and interchangeable dance tokens, enabling the synthesis of novel high-fidelity motions through the recombination of existing tokens. Due to the relative independence between upper-body and lower-body dance movements, the upper body primarily conveys emotion and stylistic details, while the lower body focuses on rhythm execution and spatial transitions. We create separate codebooks $\mathcal{Z}=\{ \mathcal{Z}^u, \mathcal{Z}^l \}$ for the upper and lower body. This decomposition also allows the combination of different code pairs to cover a wider array of choreographic units.
 
 The architecture of High-Fidelity Dance Tokenization (HFDT) stage is illustrated in Fig. \ref{fig:overview}. HFDT initiates with a Dance Encoder $\mathbf{E}$ (a three-layer 1D-CNN for information aggregation and a two-layer MLP for dimension adjustment) encoding the dance sequence $D=\{D^u, D^l\}$ into context-aware features $\mathbf{z}=\{ \mathbf{z^u}, \mathbf{z^l} \}$. These features are quantized using Finite Scalar Quantization (FSQ) to obtain $\hat{\mathbf{z}}=\{ \hat{\mathbf{z^u}}, \hat{\mathbf{z^l}} \}$, which are then decoded by Dance Decoder $\mathbf{D}$ (a two-layer MLP for dimension adjustment and a three-layer 1D TransConv for information restoration) to reconstruct the dance movement $\hat{D}=\{ \hat{D^u}, \hat{D^l} \}$. FSQ enables balanced utilization and stable gradient propagation via differentiable bounded rounding:
\begin{equation}
\hat{\mathbf{z}} = f(\mathbf{z}) + \text{sg}\left[\text{Round}[f(\mathbf{z})] - f(\mathbf{z})\right],
\end{equation}
where $f(\cdot)$ is the bounding function, set as the $\mathrm{sigmoid}(\cdot)$ function in practice. Each channel in $\hat{\mathbf{z}}$ will be quantized into one of the unique $L$ integers, therefore we have $\hat{\mathbf{z}} \in \{1,\dots,L\}^d$. The codebook size $k$ is calculated as $k = \prod_{i=1}^{d}L_i$, and $L, d$ are hyperparameter. $\text{sg}$ refers to Stop-Gradient.

\subsubsection{High-Fidelity Motion Reconstruction} Unlike VQ-VAE requiring additional loss to update any extra lookup codebook, FSQ directly integrates numerical approximations "Round" within its workflow. The Dance encoder  $\mathbf{E}$ and decoder  $\mathbf{D}$ are trained jointly via the motion reconstruction loss $\mathcal{L}_{rec}$:
\begin{equation}
\mathcal{L}_{rec} = \mathcal{L}_{\text{kin}} + \mathcal{L}_{\text{dyn}}.
\end{equation}
Simple reconstruction on SMPL parameters treats all joints equally, neglecting the complex hierarchical tree structure of human body joints, different joints vary in their tolerance to errors. For instance, errors at the root node propagate throughout all nodes, whereas errors at the hand node primarily affect only itself. Thus, we execute Forward Kinematic (FK)\cite{loper2023smpl} techniques to derive 3D joints and apply spatial reconstruction loss $\mathcal{L}_{\text{kin}}$:
\begin{equation}
    \mathcal{L}_{\text{kin}} =  \| \hat{D} - D \|_1 + \| FK(\hat{D}) - FK(D) \|_1.
\end{equation}
Moreover, we introduce the temporal loss $\mathcal{L}_{\text{dyn}}$ to better model the temporal dynamics of human motion:

\begin{equation}
\mathcal{L}_{\text{dyn}} = \alpha_1 \| \hat{D}' - D' \|_1 + \alpha_2 \| \hat{D}'' - D'' \|_1,
\end{equation}
where $D'$ and $D''$ represent the ground-truth velocity and acceleration, $\hat{D}'$ and $\hat{D}''$ are the corresponding predictions, and $\alpha_1$, $\alpha_2$ are weighting factors.

\subsection{Hybrid Music-to-Dance Generation}
Mainstream Transformer-only generation architectures \cite{siyao2022bailando,li2021ai} rely heavily on positional encodings to model temporal structure, which provides weak inductive biases for the inherent continuity in music and dance. To address this, we propose the Mamba-Transformer hybrid architecture that combines Mamba's efficient modeling of local dependencies with the Transformer's capacity for global context, thereby enhancing the aesthetic quality of generated dance.

\subsubsection{Model Architecture}
The Hybrid Music-to-Dance Generation (HMDG) stage adopts a Mamba-Transformer hybrid architecture to generate the appropriate probability distribution over dance tokens $a_{0:T'-1}=\{a_{0:T'-1}^l, a_{0:T'-1}^u\}$ given the input music $m_{1:T'}$ and genre label $g$. As illustrated in Fig.~\ref{fig:overview}, HMDG consists of three major components: the Music Encoder, Genre Encoder, and Dance Decoder. (1) \textbf{Genre Encoder.} The one-hot genre label $g$ is embedded into a learnable feature, refined via a feed-forward network, and passed through Genre Dropout to support both genre-conditioned and genre-agnostic generation. The resulting feature is then passed to the Dance Decoder via Cross-Genre SWA. (2) \textbf{Music Encoder.} Music features $m_{1:T'}$ are first extracted by the Music Foundation Model MuQ, followed by an MLP for dimension adjustment. These features are then processed by an $N_e$-layer temporal module comprising Mamba, Self-Music SWA, and Feed-Forward submodules. The resulting feature is also then passed to the Dance Decoder via Cross-Music SWA. (3) \textbf{Dance Decoder.} Previously generated dance tokens $p_{0:T'-1}=\{p_{0:T'-1}^l, p_{0:T'-1}^u\}$ are embedded, and their lower and upper features are fused by element-wise addition. The fused representation is then fed into an $N_d$-layer hierarchical multimodal processing module that integrates Mamba, Self-Dance SWA, Cross-Music SWA, Cross-Genre SWA, and Feed-Forward submodules. Finally, a linear projection is applied to predict the probability distribution over dance tokens $a_{0:T'-1}=\{a_{0:T'-1}^l, a_{0:T'-1}^u\}$ at the next timestep. During training, we apply a supervised cross-entropy loss~\cite{siyao2022bailando} to align the predicted action $a_t$ with the next-step target token $p_{t+1}$. During inference, HMDG supports: (1) autoregressive generation for short sequences ($\leq$12s); and (2) sliding-window prediction with 12s overlap for long sequences.

\subsubsection{MuQ-based Music Representation}
A powerful music representation forms the foundation of effective dance generation. Directly using raw audio is impractical due to its high temporal resolution (e.g., 16kHz), which introduces redundancy and hinders alignment with motion sequences. Existing methods either overemphasize low-level acoustic features —e.g., Librosa-based descriptors~\cite{li2021ai}—or rely solely on high-level semantic embeddings, e.g., MERT~\cite{yang2024codancers} and Jukebox~\cite{tseng2023edge}, thus failing to capture a comprehensive understanding of music relevant to dance generation. In contrast, MuQ~\cite{zhu2025muq} is a Music Foundation Model pretrained via self-supervised learning across hierarchical music informatics tasks, including beat detection, instrument classification, music tagging, and etc. MuQ not only achieves state-of-the-art performance on various benchmarks, but also adopts a lightweight architecture that supports real-time feature extraction. In practice, the extracted MuQ features are further downsampled to match the frame rate of the dance token sequence.

\subsubsection{Global-Context Modeling}
In Transformers~\cite{vaswani2017attention}, the attention layer defines computational dependencies among sequential elements and is implemented as:
\begin{equation}
\text{Attention}(Q, K, V, M) = \text{softmax}\left(\frac{QK^T + M}{\sqrt{C}}\right)V,
\end{equation}
where $Q$, $K$, and $V$ denote the query, key, and value matrices, and $M$ is the attention mask. Although music-to-dance generation is typically applied to long sequences, training is commonly conducted on short clips due to limited computational resources. During inference, the sequence is first autoregressively extended up to the training length (step 1), and then completed using a sliding window approach for the remaining portion (step 2). However, standard causal attention~\cite{radford2018improving} aligns only with step 1 and fails to model the dominant second phase during inference, resulting in a mismatch between training and inference. To mitigate this misalignment, we introduce the Sliding Window Attention (SWA) mechanism by equipping $M$ with a windowed mask that reflects the actual inference procedure.

\subsubsection{Local-Denpendency Modeling}
While the Transformer excels at long-range modeling, its position-invariant design and reliance on positional encoding~\cite{vaswani2017attention} limit its ability to capture local temporal dependencies—crucial in music and dance due to their strong local continuity. In contrast, Mamba~\cite{gu2023mamba} exhibits strong performance in fine-grained local modeling, benefiting from its sequential inductive bias~\cite{xu2024mambatalk}. We adopt Mamba to capture local dependencies via its selective state-space mechanism. Specifically, Mamba adaptively learns transition parameters through fully-connected layers and employs structured matrices to improve efficiency. At each time step $t$, the hidden state $h_t$ is updated as:
\begin{equation}
\begin{aligned}
h_t &= \bar{A}_t h_{t-1} + \bar{B}_t x_t, \\
y_t &= C_t h_t,
\end{aligned}
\end{equation}
where $\bar{A}_t, \bar{B}_t, C_t$ are dynamically updated parameters. Through discretization with sampling interval $\Delta$, the state transitions become:
\begin{equation}
\begin{aligned}
\bar{A} &= \exp(\Delta A), \\
\bar{B} &= (\Delta A)^{-1} (\exp(\Delta A) - I) \cdot \Delta B, \\
h_t &= \bar{A} h_{t-1} + \bar{B} x_t,
\end{aligned}
\end{equation}
where $(\Delta A)^{-1}$ is the inverse of $\Delta A$, and $I$ denotes the identity matrix. The scan module captures temporal dependencies by applying trainable parameters across input segments.

\begin{table*}[t]
\centering
\renewcommand{\arraystretch}{1.2}
\setlength{\tabcolsep}{5pt}
\caption{Comparison with SOTAs on the FineDance dataset.}
\small
\label{tab: comparison finedance}
\resizebox{\linewidth}{!}{
\begin{tabular}{l|cc|cc|c|ccc|cc}
\toprule
 & \multicolumn{2}{c|}{\textbf{Quality}} & \multicolumn{2}{c|}{\textbf{Synchronization}} & \textbf{Creativity} & \multicolumn{3}{|c|}{\textbf{User Study}} & \multicolumn{2}{c}{\textbf{Complexity}} \\
\cmidrule(l{-0.5pt}r{0pt}){2-3} \cmidrule(l{0pt}r{-0.5pt}){4-5} \cmidrule(l{-0.5pt}r{-0.5pt}){6-6} \cmidrule(l{0pt}r{0pt}){7-9} \cmidrule(l{0pt}r{0pt}){10-11}
 & R@5 $\uparrow$ & MM-Dist$\downarrow$ & FID$\downarrow$ & M-Dist $\downarrow$ & DIV$\uparrow$ & DS$\uparrow$ & DQ$\uparrow$ & DD$\uparrow$ & Params$\downarrow$ & Latency$\downarrow$ \\
\midrule
GT         & 23.84 & 17.77 & 0.00   & 0.00   & 17.65 & 4.6 & 4.5 & 4.5 & -    & -    \\
Random     & 3.31  & 20.63 & 402.65 & 20.25 & 1.08  & -   & -   & -   & -    & -    \\
FineNet    & 14.49 & 17.97 & 171.39 & 16.30 & 13.43 & 3.9 & 3.8 & 3.1 & \textbf{94M} & 3.97s \\
Bailando   & 13.91 & 18.56 & 142.73 & 16.15 & 15.22 & 3.9 & 3.7 & 3.7 & 152M & 5.46s \\
Lodge      & 21.85 & 17.56 & 63.28  & \textbf{11.79} & 15.24 & 4.2 & 3.8 & 3.6 & 235M & 4.57s \\
\textbf{MATHDance} & \textbf{25.83} & \textbf{17.32} & \textbf{50.81} & 12.89 & \textbf{15.47} & \textbf{4.3} & \textbf{4.2} & \textbf{4.0} & \underline{102M} & \textbf{2.97s} \\
\bottomrule
\end{tabular}
}
\end{table*}

\begin{figure}[t]
  \centering
  \includegraphics[width=\linewidth]{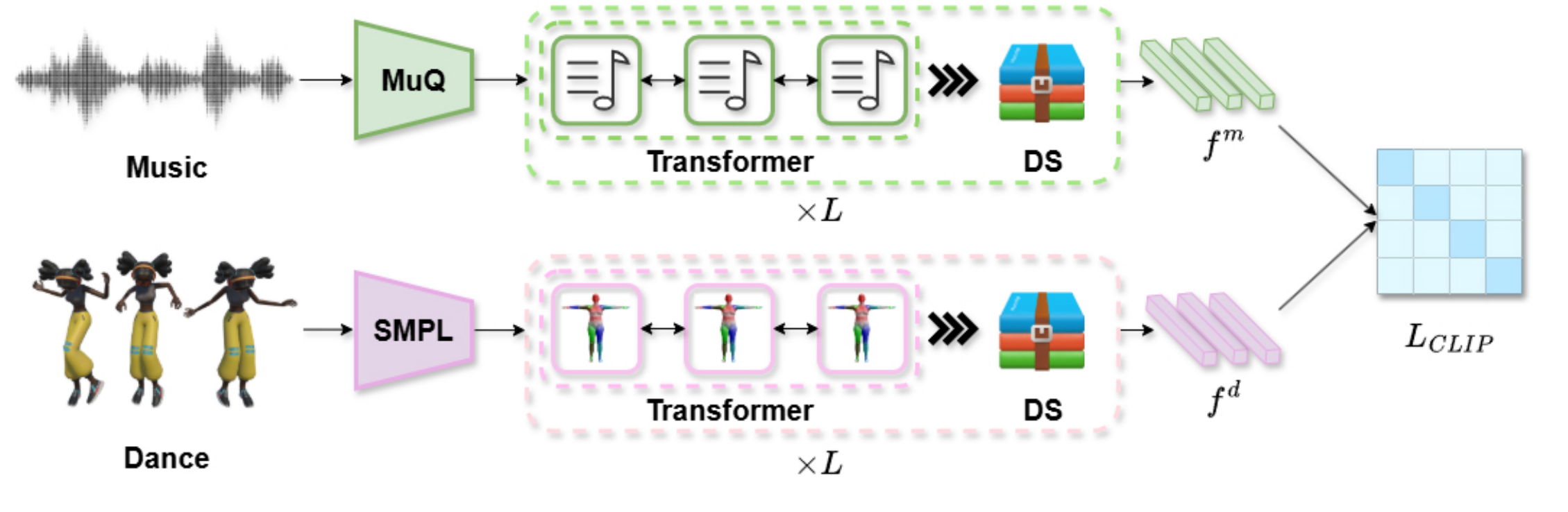}
  \caption{Architecture of the retrieval model.}
  \label{fig:Evaluation}
\end{figure}

\section{Experiment}
\subsection{Dataset}
We evaluate our method on two benchmark datasets: (1) \textbf{FineDance.} FineDance~\cite{li2023finedance} is the largest public dataset for 3D music-to-dance generation, featuring professionally performed dances captured via optical motion capture. It provides 7.7 hours of motion data at 30 fps across 16 distinct dance genres. Following~\cite{li2024lodge}, we evaluate on 20 test-set music clips, generating 1024-frame (34.13s) dance sequences. \textbf{(2) AIST++.} AIST++~\cite{li2021ai} is a widely used benchmark comprising 5.2 hours of 60 fps street dance motion, covering 10 dance genres. Following~\cite{li2021ai}, we use 40 test-set music clips to generate 1200-frame (20.00s) sequences.

\subsection{Evaluation}
Due to the subjectivity and abstraction of dance, evaluating music-to-dance generation remains a fundamental challenge. Mainstream methods lack a deep understanding of dance semantics, typically compute Stylistic, Kinetic and Geometric Features (S\&K\&G) \cite{li2021ai}-based FID for assessing dance quality and use BAS\cite{li2021ai} for synchronization. However, S\&K\&G captures only low-level motion cues, ignoring higher-level semantics. BAS, in particular, focuses narrowly on beat alignment, overlooking the multifaceted interplay between music and dance, including rhythm and expressive semantics. Given the strong coupling between movement and music, training encoders on real music-dance pairs via contrastive learning enables modeling of their shared artistic characteristics. Thus, retrieval-based protocol offer a promising alternative for evaluating generative models, and have shown strong effectiveness across domains such as text-to-motion\cite{guo2022generating}, video-to-music\cite{zhuo2023video}, and text-to-video generation\cite{liu2024evalcrafter}.
 
\subsubsection{Model Architecture}
Inspired by~\cite{yang2024beatdance}, we design a music-dance retrieval model composed of three main components: a Music Encoder, a Dance Encoder, and a Contrastive Learning module, as illustrated in Fig.~\ref{fig:Evaluation}. Our retrieval model also incorporates MuQ~\cite{zhu2025muq} and SMPL~\cite{loper2023smpl} to represent music and dance, respectively. The extracted features are subsequently fed into a multi-layer block, which consists of a Transformer-based Temporal Processing module and an Average-Pooling-based Temporal Downsampling module. Finally, the music features \( f^m \) and dance features \( f^d \) utilize CLIP~\cite{radford2021learning} loss for contrastive learning.


\subsubsection{Metric Construction}
Upon completion of model training, we extract music features $f_{m}$ and dance features $f_{d}$ using our retrieval model, both represented in a unified embedding space optimized via contrastive learning. To comprehensively assess the quality and alignment of music-to-dance generation, we introduce a suite of standardized metrics derived from these features, inspired by established practices in text-to-motion generation \cite{guo2022generating, guo2022tm2t}. (1) \textbf{Recall at 5 (R@5):} assesses macro-level semantic alignment between music and dance sequence. For evaluating music-dance alignment, we measure Recall at 5 (R@5), defined as the proportion of cases where the ground-truth music $f_{m}$ is successfully retrieved within the top-5 ranks when using generated dance features $f_{d}$ as queries. (2) \textbf{Multi-Modality Distance (MM-Dist):} evaluates micro-level feature distances between music and dances. To complement this with a fine-grained measure, we introduce Multi-Modality Distance (MM-Dist), which computes the average Euclidean distance between $f_{m}$ and $f_{d}$ across the dataset. (3) \textbf{Fréchet Inception Distance (FID):} quantifies macro-level distributional discrepancies between ground-truth and generated dances. For evaluating the global distributional similarity between generated dances and real dances, we compute the Fr'{e}chet Inception Distance (FID) on $f_{d}$, reflecting discrepancies in the overall feature space. (4) \textbf{Modality Distance (M-Dist):} measures micro-level feature distances between ground-truth and generated dances. To assess fine-grained motion fidelity, we propose the Modality Distance (M-Dist), calculated as the average Euclidean distance between generated features $f_{d}^{gen}$ and their corresponding ground-truth features $f_{d}^{gt}$ at the clip level. (5) \textbf{Diversity (DIV):} captures the creativity and variability of generated dances. Finally, to quantify the expressive richness and variability of the generated dances, we include Diversity (Div), measured as the average pairwise Euclidean distance among a batch of generated $f_{d}$ features. Together, these metrics provide a rigorous and multi-perspective evaluation framework that captures both the fidelity and semantic coherence of music-to-dance generation.



\begin{table}[t]
  \centering
  \caption{Comparison on the AIST++ dataset.}
  \label{tab: comparison aist++}
  \small
  \resizebox{\columnwidth}{!}{ 
  \begin{tabular}{@{}lccccc@{}} 
    \toprule
    Methods & R@5 $\uparrow$ & MM-Dist $\downarrow$ & FID $\downarrow$ & M-Dist $\downarrow$ & DIV $\uparrow$ \\
    \midrule
    FACT        & 10.63 & 20.78 & 126.44 & 16.81 & 12.37 \\
    Bailando    & 16.87 & 18.71 & 92.77  & 14.38 & 14.89 \\
    EDGE        & 18.45 & 18.93 & 86.03  & 13.92 & 14.03 \\
    Lodge       & 21.66 & 17.24 & 67.81  & 13.62 & 15.21 \\
    \textbf{Ours}        & \textbf{22.74} & \textbf{17.01} & \textbf{50.82}  & \textbf{12.45} & \textbf{15.35} \\
    \bottomrule
  \end{tabular}}
\end{table}

\begin{table}[t]
  \centering
  \caption{Ablation on Music-to-Dance Generation stage.}
  \label{tab: ablation music-to-dance generation}
  \small
  \resizebox{\columnwidth}{!}{ 
  \begin{tabular}{@{}lccccc@{}} %
        \toprule
    Methods & R@5 $\uparrow$ \ & MM-Dist $\downarrow$ & FID $\downarrow$ & M-Dist $\downarrow$ & DIV $\uparrow$ \\
    \midrule
      GT          & 23.84 & 17.77 &   0.00 &  0.00 & 17.65 \\
    Random      &  3.31 & 20.63 & 402.65 & 20.25 &  1.08 \\
    w/o SWA     & 15.89 & 18.04 & 104.19 & 13.93 & 14.63 \\
    w/o MuQ     & 21.86 & 18.66 & 63.27 & 15.34 & \textbf{16.40} \\
    w/o Mamba   & 17.21 & 18.35 &  80.27 & 15.92 & 16.21 \\
    Ours        & \textbf{25.83} & \textbf{17.32} & \textbf{50.81} & \textbf{12.89} & 15.47 \\
    \bottomrule
  \end{tabular}}
\end{table}

\subsection{Comparison}

\subsubsection{Quantitative Analysis}
To compare the performance of different methods in terms of  quantitative analysis, we evaluate MATHDance against Bailando\cite{siyao2022bailando}, FineNet\cite{li2023finedance} and Lodge\cite{li2024lodge}. To provide proper reference, we also report evaluation metrics on real dance (GT) and randomly Gaussian noise (Random). As shown in Tab.~\ref{tab: comparison aist++} and Tab.~\ref{tab: comparison finedance}, MATHDance achieves superior performance across all metrics on various dataset. On FineDance, it improves R@5 by 3.98 and reduces FID by 12.47 compared to Lodge, while maintaining strong MM-Dist and M-Dist scores. On AIST++, it further improves R@5 by 1.08 and reduces FID by 16.99, showing robustness across datasets. DIV remains competitive in both settings. In summary, by addressing the challenge of choreographic consistency, MATHDance achieves significant improvements in both dance quality and dance synchronization.

\subsubsection{User Study}
Dance’s inherent subjectivity makes user feedback essential for evaluating generated movements\cite{legrand2009perceiving}, particularly in the music-to-dance generation. We select 30 music segments (34.13 seconds each) and generate dance sequence using models mentioned above. These sequences are evaluated through a double-blind questionnaire, by 30 participants with backgrounds in dance practice. The questionnaires are based on a 5-point scale (Great, Good, Fair, Bad, Terrible) and assess three aspects: Dance Synchronization (DS, alignment with rhythm and style), Dance Quality (DQ, physical plausibility and aesthetics), and Dance Diversity (DD, variety and creativity). As shown in Tab. \ref{tab: comparison finedance}, MATHDance significantly outperforms the other methods across all metrics (DS = 4.3, DQ = 4.2, DD = 4.0), demonstrating its superiority in terms of human preferences.

\subsubsection{Complexity Analysis}
We also include a comparison of generation complexity for producing a 1024-frame (34.13s) dance sequence. All latency evaluations are conducted on an RTX 3090 GPU with an Intel Xeon Gold 5218 CPU. As shown in Tab.~\ref{tab: comparison finedance}, MATHDance demonstrates clear superiority in generation efficiency. It has a comparable parameter size to FineNet (102M vs. 94M), and achieves the lowest inference latency of 2.97s, outperforming all baselines by a large margin. This speed enables seamless integration into interactive systems, where rapid feedback is crucial for user engagement in practice.

\begin{table}[t]
  \centering
  \caption{Ablation on Dance Tokenization stage.}
  \resizebox{\columnwidth}{!}{%
  \begin{tabular}{l|cc|cc|ccc}
    \toprule
\multirow{2}{*}{\makecell{\centering Model}} 
 & \multicolumn{2}{c|}{Joints} & \multicolumn{2}{c|}{SMPL} & \multicolumn{3}{c}{CUR} \\
    \cmidrule{2-8}
    & MSE $\downarrow$ & MAE $\downarrow$ & MSE $\downarrow$ & MAE $\downarrow$ & T@1 $\uparrow$ & T@5 $\uparrow$ & T@10 $\uparrow$ \\
    \midrule
    FSQ (Comp.+Spat.+Temp.) & \textbf{0.0076} & \textbf{0.0491} & \textbf{0.0238} & \textbf{0.0847} & 100\% & 100\% & 100\% \\
    FSQ (Comp.+Spat.)       & 0.0089 & 0.0507 & 0.0240 & 0.0859 & 100\% & 100\% & 100\% \\
    \midrule
    \makecell[l]{FSQ (Comp.)} & 0.0123 & 0.0606 & 0.0253 & 0.0866 & 100\% & 100\% & 100\% \\
    \makecell[l]{VQ-VAE (Comp.)} & 0.0168 & 0.0737 & 0.0349 & 0.1038 & 98.80\% & 63.05\% & 37.60\% \\
    \makecell[l]{FSQ (Res.)} & 0.0159 & 0.0714 & 0.0258 & 0.0909 & 87.35\% & 79.65\% & 73.90\% \\
    \makecell[l]{VQ-VAE (Res.)} & 0.0220 & 0.0842 & 0.0308 & 0.0984 & 71.90\% & 46.75\% & 31.85\% \\
    \makecell[l]{FSQ (Orig.)} & 0.0182 & 0.0757 & 0.0257 & 0.0912 & 82.24\% & 23.44\% & 6.67\% \\
    \makecell[l]{VQ-VAE (Orig.)} & 0.0204 & 0.0807 & 0.0280 & 0.0932 & 47.29\% & 14.67\% & 5.86\% \\
    \bottomrule
  \end{tabular}%
  }
  \label{tab: ablation study dance reconstruction}
\end{table}

\begin{table}[t]
  \centering
    \caption{Ablation for dance-to-music retrieval.}
  \label{tab: ablation study retrieval}
  \resizebox{\columnwidth}{!}{ 
  \begin{tabular}{lcccc}
    \toprule
    Model & Recall@5 $\uparrow$ & Recall@10 $\uparrow$ & Median Rank $\downarrow$ & Mean Rank $\downarrow$ \\
    \midrule
    AVG->CNN & 15.23 & 31.79 & 19.0 & 25.42 \\
    MuQ->Librosa & 11.26 & 23.84 & 25.0 & 36.26 \\
    Transformer->Mamba & 5.96 & 14.57 & 34.0 & 45.22 \\
    Transformer+Mamba & 22.51 & \textbf{38.41}  & 15.0 & 23.54 \\
    Ours & \textbf{23.84} & \textbf{38.41} & \textbf{14.0} & \textbf{18.79} \\
    \bottomrule
  \end{tabular}}
\end{table}

\begin{figure}[t]
  \centering
  \includegraphics[width=\linewidth]{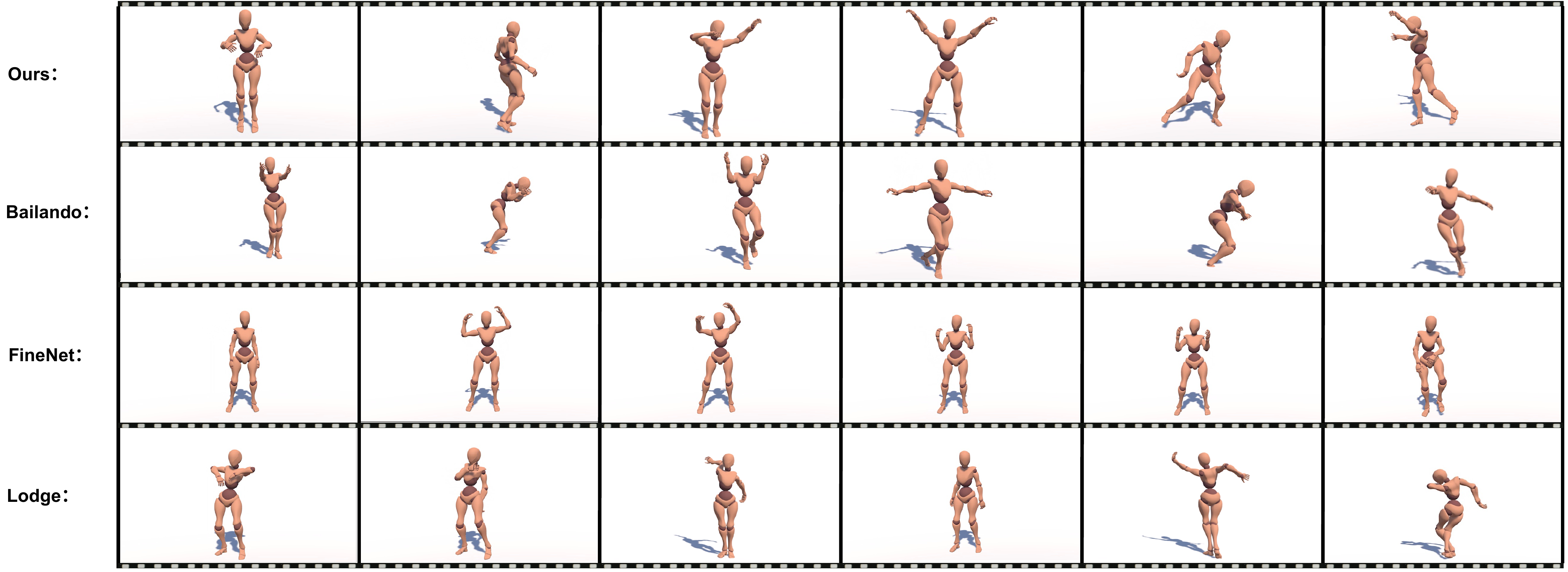}
    \vspace{-0.2in}
  \caption{Comparison with SOTAs on a soft folk music clip.}
  \label{fig: comparison}
    \vspace{-0.1in}
\end{figure}

\begin{figure}[t]
  \centering
  \includegraphics[width=\linewidth]{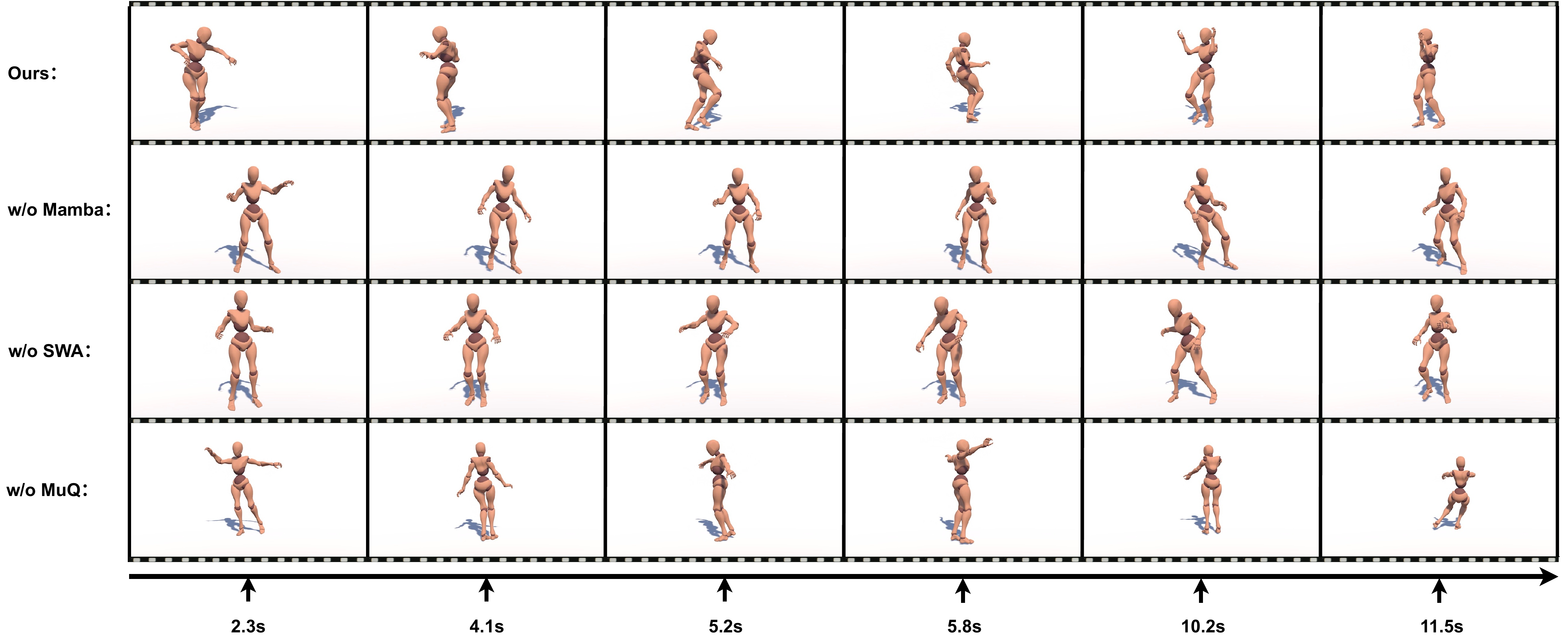}
  \vspace{-0.2in}
  \caption{Ablation on Music-to-Dance Generation stage.}
  \label{fig: ablation HMDGS}
  \vspace{-0.1in}
\end{figure}

\subsection{Ablation Study}
\subsubsection{Hybrid Music-to-Dance Generation Stage}
In this section, we investigate the impacts of Sliding Window Attention (SWA), MuQ-based Music Representation (MuQ), and Mamba-enhanced Architecture (Mamba), as detailed in Tab.~\ref{tab: ablation music-to-dance generation}.  \textbf{(1) SWA}. The SWA is utilized to adapt to the sliding-window inference phase, which is prevalent in long-sequence generation. Replacing sliding window attention with causal attention results in a notable decrease across all metrics, affirming SWA's critical role. \textbf{(2) MuQ}. To enhance music representation, we integrate MuQ, which, when replaced with Librosa, leads to obvious declines in R@5/MM-Dist/FID/M-Dist, while DIV shows a slight increase, thus validating MuQ’s effectiveness in driving dance generation. \textbf{(3) Mamba}. The Mamba is introduced to bolster local dependencies, where significant improvements are noted in R@5/MM-Dist (8.52/1.03) and FID/M-Dist (29.46/3.03), showcasing its capability to enhance music-dance synchronization and dance quality.

\subsubsection{High-Fidelity Dance Tokenization Stage}
In this section, we investigate the impacts of Quantization Strategy, Spatial Loss, and Temporal Loss, as detailed in Tab.~\ref{tab: ablation study dance reconstruction}. Here we compare the performance of methods: Original Motion FSQ (Orig.), Original Motion VQ-VAE (Orig.), Residual Motion FSQ (Res.), Residual Motion VQ-VAE (Res.), Compositional Motion FSQ (Comp.), Compositional Motion VQ-VAE (Comp.), Compositional Motion FSQ with Spatial Loss (Comp.+Spat.) and Compositional Motion FSQ with Spatial and Temporal Loss (Comp.+Spat.+Temp.). To ensure fairness, we set the codebook number to 2/2/1 and the codebook size to 1024/1024/65536 for Comp., Res., and Orig., respectively. We evaluate by MSE/MAE on SMPL and 3D Joints, and analyze the Codebook Utilization Rate (CUR), which measures the proportion of codebook entries whose usage exceeds thresholds of 1, 5, and 10 times. Additionally, the codebook was used 192,375 times in total. \textbf{(1) Quantization Strategy.} To mitigate codebook collapse in VQ-VAE, we adopt FSQ with differentiable scalar quantization replacing discrete argmin selection. Across all codebook structures—Original, Compositional, and Residual—FSQ consistently outperforms VQ-VAE in SMPL and 3D joint reconstruction. Moreover, the superior result on CUR:T@1/5/10 indicates its efficient codebook utilization. \textbf{(2) Spatial Loss.} To enhance the spatial naturalness, we introduce Spatial Loss by imposing constraints on 3D joints derived from Forward Kinematics (FK), which yields significant improvements in both SMPL and joint reconstruction. \textbf{(3) Temporal Loss.} To improve temporal fidelity, we apply Temporal Loss by constraining joint velocity and acceleration, which also brings notable gains in reconstruction accuracy.

\subsubsection{Ablation Study for Retrieval Model}
Leveraging the efficiency and lower complexity inherent in retrieval tasks, we employ retrieval models to evaluate generative models. We investigate the effects of Temporal Downsampling, Music Representation, and Temporal Modeling in our proposed dance-music retrieval model, using Recall@5/10 and Median/Mean Rank as evaluation metrics. As shown in Tab. \ref{tab: ablation study retrieval}, replacing average pooling (AVG) with CNN for temporal downsampling leads to a noticeable drop across all metrics. Replacing MuQ with Librosa results in an even larger performance decline, also demonstrating MuQ’s strong capability in music feature representation. Moreover, neither replacing nor adding Mamba layers brings improvement, suggesting that Mamba is more suitable for music-to-dance generation tasks with strong local dependency, rather than retrieval tasks that require global semantic understanding.

\begin{figure}[t]
  \centering
  \includegraphics[width=\linewidth]{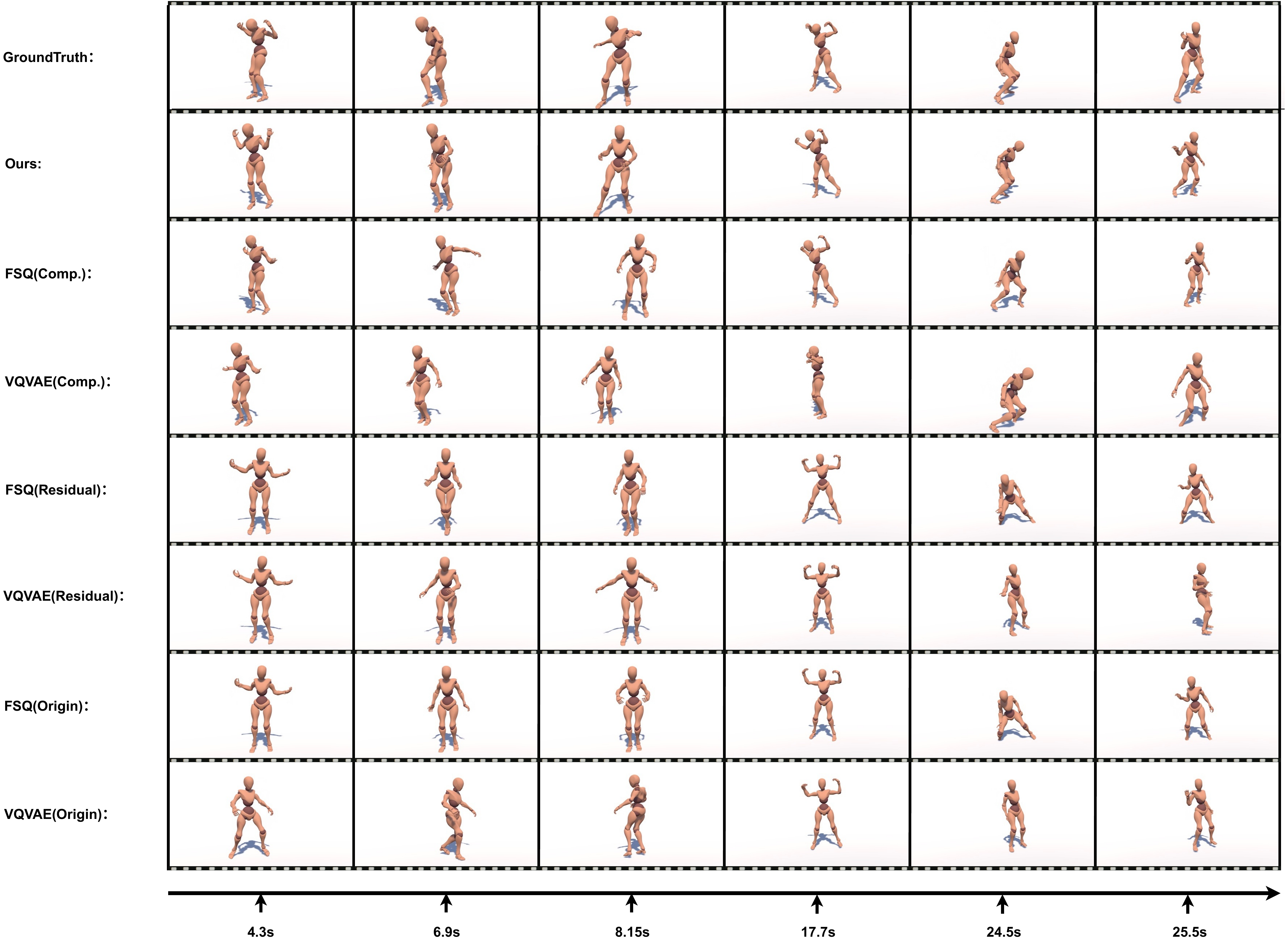}
  \vspace{-0.2in}
  \caption{Ablation on Dance Tokenization stage.}
  \label{fig: ablation KDQS}
  \vspace{-0.2in}
\end{figure}

\subsection{Qualitative Analysis}
\subsubsection{Comparison for Music-to-Dance Generation}
To assess the visual quality of the generated dance sequences, we perform a qualitative comparison between MATHDance and several existing baseline models, as depicted in Fig. \ref{fig: comparison}. In terms of expressiveness, MATHDance outperforms the competing methods in several key areas. Specifically, compared to MATHDance, Lodge occasionally generates awkward poses, such as unnatural limb bending; FineNet often exhibits excessive repetition of motion patterns; Bailando lacks expressiveness and visual aesthetics. These findings underscore the superiority of MATHDance in generating dance with physical plausibility and aesthetic quality.

\subsubsection{Ablation on Dance Generation stage}
In this section, we explore the impact of all components in the Hybrid Music-to-Dance Generation Stage from a qualitative perspective. As shown in Fig.~\ref{fig: ablation HMDGS}, Slide Window Attention (SWA), MuQ-based Music Representation (MuQ), and Mamba-enhanced temporal modeling (Mamba) each demonstrate their crucial role. Specifically, removing Mamba results in simpler, repetitive movements with weak spatial dynamics and rhythm alignment. Removing SWA leads to minimal, low-energy motion with almost no displacement or musical interaction. Removing MuQ causes a mismatch with strong percussion beats, reducing the coherence between music and motion. In conclusion, all components in this stage effectively contribute to the overall performance.

\subsubsection{Ablation on Dance Tokenization stage}
In this section, we conduct a qualitative analysis to compare our method with Original Motion FSQ (Orig.), Original Motion VQ-VAE (Orig.), Residual Motion FSQ (Res.), Residual Motion VQ-VAE (Res.), Compositional Motion FSQ (Comp.), and Compositional Motion VQ-VAE (Comp.). As shown in Fig.~\ref{fig: ablation KDQS}, our method achieves the most accurate dance reconstruction results. Under the same codebook structure, FSQ-based quantization strategies clearly outperform those based on VQ-VAE. Under identical quantization settings, the compositional structure exhibits superior performance compared to residual and original structures. Furthermore, our method, enhanced by spatial-temporal constraints, surpasses FSQ (Comp.). These results collectively demonstrate the superiority of our approach in the High-Fidelity Dance Tokenization stage.

\begin{table}[t]
  \centering
  \caption{Exploring the reliability of evaluation protocol.}
  \label{tab: user study evaluation}
  \resizebox{\columnwidth}{!}{ 
  \begin{tabular*}{\columnwidth}{@{\extracolsep{\fill}}lccc}
    \toprule
    \multicolumn{1}{c}{Evaluation} & 
    \multicolumn{1}{c}{DS $\uparrow$} & 
    \multicolumn{1}{c}{DQ $\uparrow$} & 
    \multicolumn{1}{c}{DD $\uparrow$} \\
    \midrule
    BAS & 72\%  & - & - \\
    Kinetic & - & 59\% & 55\% \\
    Geometric & - & 52\% & 58\% \\
    Ours & \textbf{84\%} & \textbf{77\%} & \textbf{71\%} \\
    \bottomrule
  \end{tabular*}
  }
\end{table}

\subsection{Reliability of Evaluation Protocol}
To validate metric reliability, we assess metric-user alignment using randomly selected 100 generated dance pairs from User Study in Sec. 4.3.3. We calculate the alignment rate, defined as the proportion of pairs where human preferences and evaluation metrics consistently agree on the preferred or less preferred option. We compare our retrieval model to Kinetic and Geometric baselines by measuring alignment with DQ (via M-Dist/FID) and DD (via DIV), and assess how well DS aligns with R@5/MM-Dist versus BAS. As shown in Tab. \ref{tab: user study evaluation}, our evaluation metrics achieve strong alignment with human judgments across all three aspects: approximately 12\% higher than the second-best in DS, 18\% higher in DQ, and 13\% higher in DD. The excellent result also validates the effectiveness of using retrieval-based evaluation for generative models in the music-to-dance task.

\subsection{Implementation Details}

\subsubsection{High-Fidelity Dance Tokenization}
\noindent\textbf{Data.}
We train the tokenizer on 12\,s SMPL sequences (30\,fps), i.e., $T{=}360$ and
$X,\hat{X}\in\mathbb{R}^{360\times147}$ (6D rotations).
Training clips are constructed using a sliding window (window 360, stride 30).
\noindent\textbf{Architecture.}
Upper- and lower-body branches share the same codebook configuration.
A 3-layer CNN encoder $E$ temporally downsamples the input, and a 3-layer
transposed-conv decoder $G$ upsamples it, producing latent codes
$p^{u},p^{l}\in\mathbb{R}^{45}$ ($T'{=}45$).
We use Finite Scalar Quantization with codebook size 1000, $L{=}[8,5,5,5]$,
and feature dimension 512.
\noindent\textbf{Loss \& Optimization.}
We supervise reconstruction with an SMPL-parameter loss $\mathcal{L}_{\text{smpl}}$
and a joint-position loss $\mathcal{L}_{\text{joint}}$, augmented with velocity and
acceleration terms weighted by $\alpha_1{=}0.5$ and $\alpha_2{=}0.25$.
We train for 200 epochs with Adam ($\beta{=}(0.5,0.99)$), fixed learning rate,
and batch size 32.

\subsubsection{Hybrid Music-to-Dance Generation}
\noindent\textbf{Data.}
We train on the quantized latent codes $(p^{u},p^{l})$ extracted from 12\,s sequences,
using the same sliding-window augmentation (window 360, stride 30).
\noindent\textbf{Model.}
MATHDance adopts a Mamba--Transformer hybrid with $N_e{=}6$ layers in the music encoder
and $N_d{=}6$ layers in the dance decoder.
To handle missing genre labels at test time, we apply genre dropout (0.3) in the genre encoder.
The Mamba block uses model dim 512, state size 16, conv kernel 4, and expansion 2.
The Transformer block uses hidden size 512, 8 heads, FFN dim 2048, and dropout 0.25.
For sliding-window attention, we use an autoregressive step of 30 and window stride 15.
All submodules use LayerNorm and residual connections.

\noindent\textbf{Inputs/Outputs.}
Genre labels (16 FineDance classes) are embedded to 512 dims via \texttt{nn.Embedding}.
MuQ music features (1024 dims) are projected to 512 dims using a 2-layer MLP.
The decoder predicts a 2000-way softmax distribution: indices 0--999 correspond to the
upper-body codebook and 1000--1999 to the lower-body codebook.
\noindent\textbf{Optimization.}
We train for 300 epochs with Adam ($\beta{=}(0.9,0.99)$), fixed learning rate,
and batch size 128.

\subsubsection{Dance-Music Retrieval for Evaluation}
We train a dance--music retrieval model on SMPL motion and MuQ music representations.
Training pairs are constructed with a sliding window (window 360, stride 180).
Both music and dance encoders use $L{=}9$ Transformer layers (hidden 512, 8 heads, dropout 0.25),
with temporal downsampling implemented by average pooling (rate 2).
We optimize with CLIP loss (temperature 4.6052) using Adam
(lr $1{\times}10^{-5}$, $\beta{=}(0.5,0.999)$, batch size 32),
and apply StepLR (step 5, $\gamma{=}0.33$).

\section{Conclusion}
In this paper, we propose MATHDance, a two-stage latent-space framework for music-to-dance generation that ensures choreographic consistency through the High-Fidelity Dance Tokenization stage for physical plausibility, and the Hybrid Music-to-Dance Generation stage for aesthetic quality. Additionally, we introduce a retrieval-based evaluation protocol for music-to-dance generation. Experiments on the FineDance and AIST++ datasets demonstrate MATHDance's superiority in various aspects, and the reliability of our evaluation protocol. In future work, we plan to extend MATHDance with motion or text conditioning to enable more interactive and flexible dance generation.

\bibliographystyle{ACM-Reference-Format}
\balance
\bibliography{mm-paper}

\end{document}


\title{MatchDance: Collaborative Mamba-Transformer Architecture Matching for High-Quality 3D Dance Synthesis}

\renewcommand{\shortauthors}{Trovato et al.}

\begin{abstract}
This appendix contains additional materials for the paper “MATHDance: Mamba-Transformer Architecture with Uniform Tokenization for High-Quality 3D Dance Generation”. The appendix is organized as follows: 
    \begin{itemize}
      \item Implementation Details.
      \item Qualitative Analysis.
      \item Evaluation Metrics.
      \item  Future Work
      \item  Questionnaire
    \end{itemize}
\end{abstract}

\ccsdesc[500]{Applied computing~Performing arts}
\ccsdesc[300]{Human-centered computing~HCI design and evaluation methods}

\keywords{AI for Art, Multimedia Learning, AI Generative Content, 3D Human Motion Generation, Music-Driven Dance Generation}

\received{20 February 2007}
\received[revised]{12 March 2009}
\received[accepted]{5 June 2009}


\maketitle

\section{Implementation Details}
\subsection{High-Fidelity Dance Tokenization}
In the High-Fidelity Quantization Stage, we use a shared codebook configuration for the upper and lower body branches. The model is trained on 12-second SMPL 6D rotation sequences sampled at 30fps, where \( D, \hat{D} \in \mathbb{R}^{360 \times 147} \) (i.e., \( T = 360 \)). For data construction, we augment the training set using a sliding window approach with a window size of 360 and a stride of 30. A three-layer CNN encoder \( E \) performs temporal downsampling, and a three-layer transposed convolution decoder \( D \) performs upsampling. The latent codes for the lower and upper body are \( p^l, p^u \in \mathbb{R}^{45} \), with \( T' = 45 \). In the Finite Scalar Quantization module, the codebook size is 1000, with \( L = [8, 5, 5, 5] \), and the feature dimension is set to 512. For reconstruction, we use both SMPL-parameter loss \( \mathcal{L}_{\text{smpl}} \) and joint-position loss \( \mathcal{L}_{\text{joint}} \), with velocity and acceleration terms weighted by \( \alpha_1 = 0.5 \) and \( \alpha_2 = 0.25 \), respectively. The model is trained for 200 epochs using the Adam optimizer, with exponential decay rates of 0.5 and 0.99 for the first and second moment estimates. A fixed learning rate is used with a batch size of 32.

\subsection{Hybrid Music-to-Dance Generation}
In the Hybrid Music-to-Dance Generation Stage, we adopt a Mamba-Transformer hybrid architecture, trained on latent codes \( p^l, p^u \in \mathbb{R}^{45} \) extracted from the High-Fidelity Dance Tokenization stage, using 12-second dance sequences at 30fps. For data construction, we augment the training set using a sliding window approach with a window size of 360 and a stride of 30. In MATHDance, the Music Encoder consists of \( N_e = 6 \) processing layers, and the Dance Decoder is also composed of \( N_e = 6 \) processing layers. To simulate real-world scenarios where genre labels may be unavailable, the Genre Encoder incorporates a genre dropout of 0.3. The Mamba block is configured with a model dimension of 512, state size of 16, convolution kernel size of 4, and expansion factor of 2. The Transformer block uses a hidden size of 512, 8 attention heads, a feedforward dimension of 2048, and a dropout rate of 0.25. For Slide Window Attention, we set the autoregressive step to 30 and the sliding window stride to 15 to construct the attention matrix. All submodules in MATHDance apply layer normalization and residual addition. For input representation, genre labels (16 classes from FineDance) are embedded using \texttt{nn.Embedding} to match the 512 dimensions, while music features extracted by MuQ (1024 dimensions) are projected to 512 dimensions via a two-layer MLP. For output, MATHhDance predicts 2000-class distributions via softmax: indices 0–999 represent the upper-body codebook, and 1000–1999 represent the lower-body codebook. The model is optimized using Adam with exponential decay rates of 0.9 and 0.99 for the first and second moment estimates, respectively, trained for 300 epochs with a fixed learning rate and a batch size of 128.

\subsection{Dance-Music Retrieval for Evaluation}
To better evaluate generative models, we construct a dance-music Retrieval model trained directly on SMPL-based dance sequences and MuQ-based music representations. For data construction, we adopt a sliding window strategy with a window size of 360 and a stride of 180. Both the Music Encoder and Dance Encoder consist of \( L = 9 \) stacked temporal processing layers. Each Transformer layer has a hidden size of 512, 8 attention heads, a dropout rate of 0.25, and contains a single encoder layer. Temporal downsampling is implemented via average pooling with a downsampling rate of 2. The CLIP loss is used for training with a temperature coefficient set to 4.6052. The model is optimized using Adam with a learning rate of \( 1 \times 10^{-5} \), \(\beta = (0.5, 0.999)\), and a batch size of 32. A StepLR scheduler is applied with a step size of 5 and a decay factor of 0.33.

\section{Qualitative Analysis}
\begin{figure*}[t]
  \centering
  \includegraphics[width=0.95\linewidth]{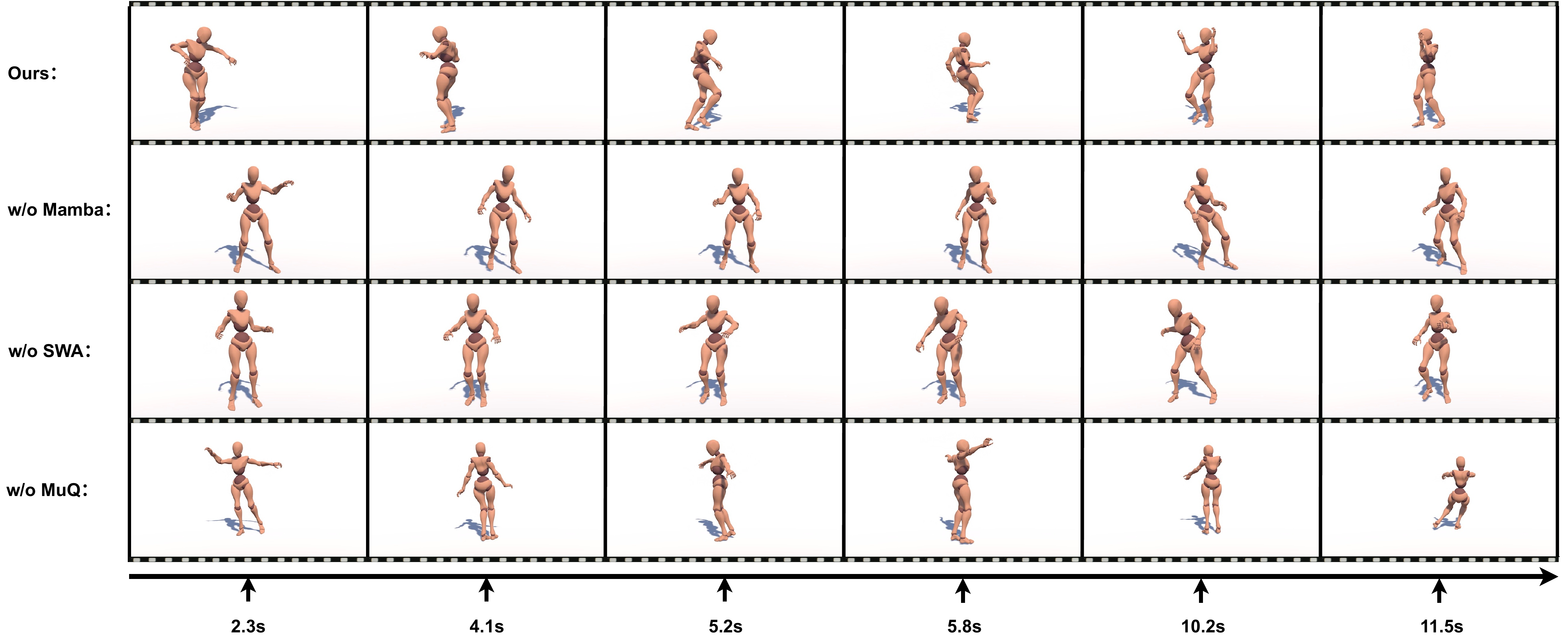}
  \caption{Visualization of Ablation Study for Hybrid Music-to-Dance Generation Stage.}
  \label{fig: ablation HMDGS}
\end{figure*}

\subsection{Hybrid Music-to-Dance Generation}
In this section, we explore the impact of all components in the Hybrid Music-to-Dance Generation Stage from a qualitative perspective. As shown in Fig.~\ref{fig: ablation HMDGS}, Slide Window Attention (SWA), MuQ-based Music Representation (MuQ), and Mamba-enhanced temporal modeling (Mamba) each demonstrate their crucial role. Specifically, removing Mamba results in simpler, repetitive movements with weak spatial dynamics and rhythm alignment. Removing SWA leads to minimal, low-energy motion with almost no displacement or musical interaction. Removing MuQ causes a mismatch with strong percussion beats, reducing the coherence between music and motion. In conclusion, all components in this stage effectively contribute to the overall performance.

\begin{figure*}[t]
  \centering
  \includegraphics[width=0.95\linewidth]{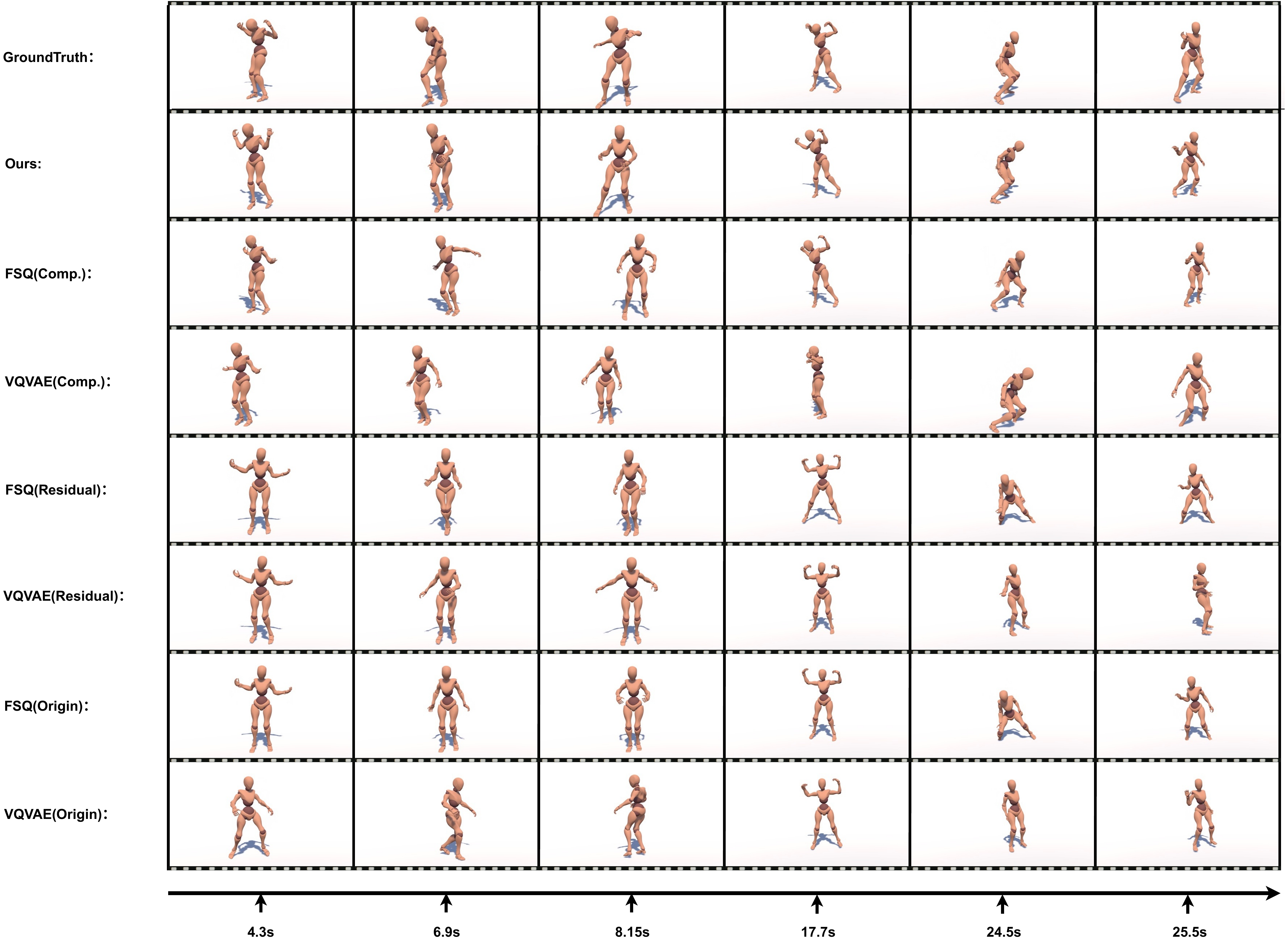}
  \caption{Visualization of Ablation Study for High-Fidelity Dance Tokenization stage.}
  \label{fig: ablation KDQS}
\end{figure*}

\subsection{High-Fidelity Dance Tokenization}
In this section, we conduct a qualitative analysis to compare our method with Original Motion FSQ (Orig.), Original Motion VQ-VAE (Orig.), Residual Motion FSQ (Res.), Residual Motion VQ-VAE (Res.), Compositional Motion FSQ (Comp.), and Compositional Motion VQ-VAE (Comp.). As shown in Fig.~\ref{fig: ablation KDQS}, our method achieves the most accurate dance reconstruction results. Under the same codebook structure, FSQ-based quantization strategies clearly outperform those based on VQ-VAE. Under identical quantization settings, the compositional structure exhibits superior performance compared to residual and original structures. Furthermore, our method, enhanced by spatial-temporal constraints, surpasses FSQ (Comp.). These results collectively demonstrate the superiority of our approach in the High-Fidelity Dance Tokenization stage.

\section{Evaluation Protocol}
Due to the subjectivity and abstraction of dance, evaluating music-to-dance generation remains a fundamental challenge. Existing methods lack a deep understanding of dance semantics, typically compute Kinetic and Geometric Feature(K\&G)\cite{li2021ai}-based FID for assessing dance quality and use BAS\cite{li2021ai} for synchronization. However, K\&G captures only low-level motion cues, ignoring higher-level semantics such as dancer emotion and dance style. BAS, in particular, focuses narrowly on beat alignment, overlooking the multifaceted interplay between music and dance, including rhythm and expressive semantics. Given the strong coupling between movement and music, training encoders on real music-dance pairs via contrastive learning enables modeling of their shared artistic characteristics. Owing to their efficiency and simplicity, retrieval-based methods offer a promising alternative for evaluating generative models, and have shown strong effectiveness across domains such as text-to-motion\cite{guo2022generating,guo2022tm2t}, video-to-music\cite{zhuo2023video}, and text-to-video generation\cite{huang2024vbench,liu2024evalcrafter}.

Upon completion of model training, we extract music features $f_{m}$ and dance features $f_{d}$ using our retrieval model, both represented in a unified embedding space optimized via contrastive learning. To comprehensively assess the quality and alignment of music-to-dance generation, we introduce a suite of standardized metrics derived from these features, inspired by established practices in text-to-motion generation \cite{guo2022generating, guo2022tm2t}. (1) \textbf{Recall at 5 (R@5):} assesses macro-level semantic alignment between music and dance sequence. For evaluating music-dance alignment, we measure Recall at 5 (R@5), defined as the proportion of cases where the ground-truth music $f_{m}$ is successfully retrieved within the top-5 ranks when using generated dance features $f_{d}$ as queries. (2) \textbf{Multi-Modality Distance (MM-Dist):} evaluates micro-level feature distances between music and dances. To complement this with a fine-grained measure, we introduce Multi-Modality Distance (MM-Dist), which computes the average Euclidean distance between $f_{m}$ and $f_{d}$ across the dataset. (3) \textbf{Fréchet Inception Distance (FID):} quantifies macro-level distributional discrepancies between ground-truth and generated dances. For evaluating the global distributional similarity between generated dances and real dances, we compute the Fr'{e}chet Inception Distance (FID) on $f_{d}$, reflecting discrepancies in the overall feature space. (4) \textbf{Modality Distance (M-Dist):} measures micro-level feature distances between ground-truth and generated dances. To assess fine-grained motion fidelity, we propose the Modality Distance (M-Dist), calculated as the average Euclidean distance between generated features $f_{d}^{gen}$ and their corresponding ground-truth features $f_{d}^{gt}$ at the clip level. (5) \textbf{Diversity (DIV):} captures the creativity and variability of generated dances. Finally, to quantify the expressive richness and variability of the generated dances, we include Diversity (Div), measured as the average pairwise Euclidean distance among a batch of generated $f_{d}$ features. Together, these metrics provide a rigorous and multi-perspective evaluation framework that captures both the fidelity and semantic coherence of music-to-dance generation.

\begin{figure}[h]
  \centering
\includegraphics[width=0.45\linewidth]{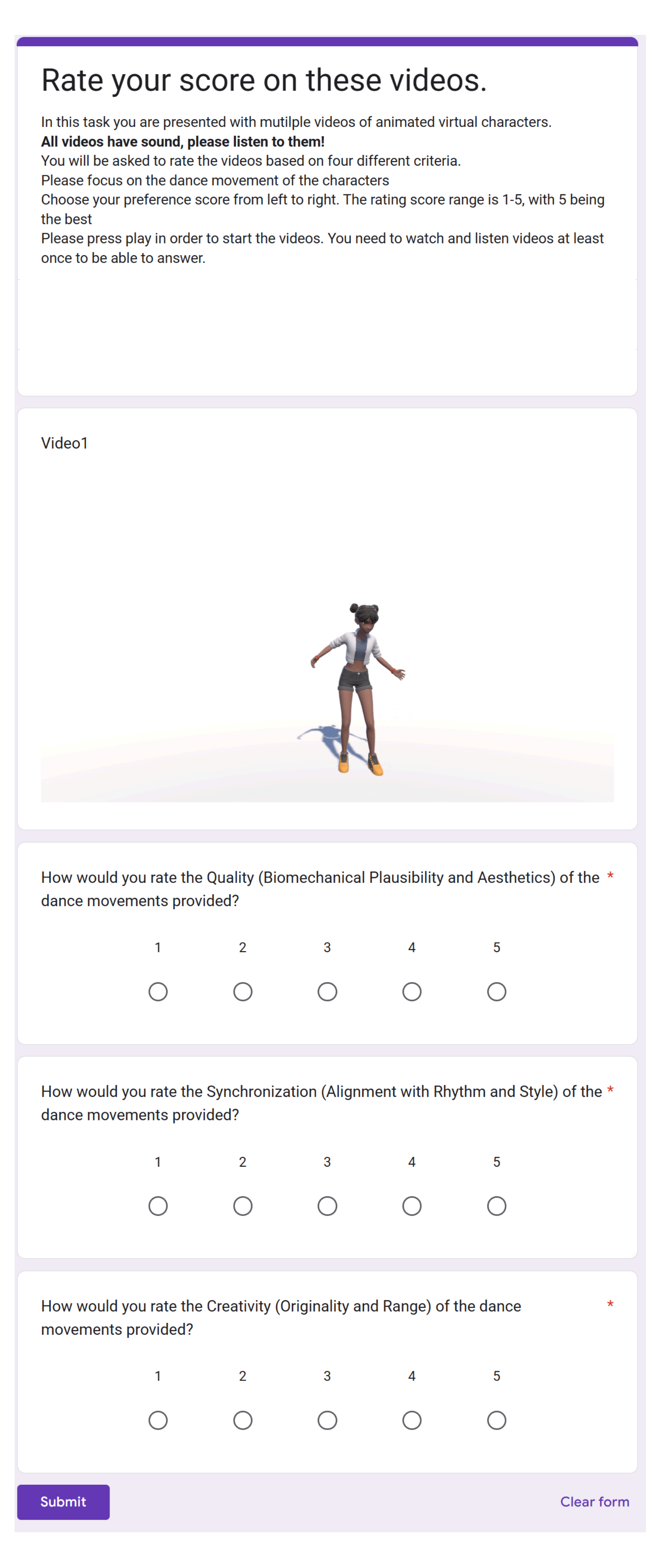}
  \caption{The screenshots of user study website.}
  \label{fig: user study questionnaire}
\end{figure}

\section{Questionnaire}
Dance’s inherent subjectivity makes user feedback essential for evaluating generated movements\cite{legrand2009perceiving}, particularly in the music-to-dance generation. We select 30 music segments (34.13 seconds each) and generate dance sequence using models mentioned above. These sequences are evaluated through a double-blind questionnaire, by 30 participants with backgrounds in dance practice. The questionnaires are based on a 5-point scale (Great, Good, Fair, Bad, Terrible) and assess three aspects: Dance Synchronization (DS, alignment with rhythm and style), Dance Quality (DQ, physical plausibility and aesthetics), and Dance Diversity (DD, variety and creativity). The screenshot of our user study website is shown in Fig. \ref{fig: user study questionnaire}, displaying the template layout presented to the participants. In addition to the main trials, participants are also subjected to several catch trials, which involved displaying Ground Truth videos and videos with distorted motion. Participants who failed to rate the GT videos higher and the distorted motion videos lower are considered unresponsive or inattentive, and their data were excluded from the final evaluation.

\section{Future Work}
\subsection{Customized Dance Generation}
Current music-to-dance generation paradigms predominantly rely on fixed music inputs, limiting user controllability and failing to meet the growing demand for personalized choreography. Current research on controllable generation remains limited, and the level of control achieved is insufficient for practical applications\cite{huang2022genre,li2024exploring,zhuang2023gtn}. Future directions may involve incorporating user-controllable conditions such as genre labels, textual descriptions, or partial pose guidance to enable more expressive and customized dance generation. This line of work holds potential for enhancing user creativity and interactivity in content-driven applications.

\subsection{Noise-Resistant Dance Generation}
3D motion capture data often suffer from noise artifacts such as sudden positional jumps or temporal discontinuities, as observed even in high-quality datasets like FineDance. Moreover, the limited scale of 3D dance datasets makes models prone to overfitting. Future research should explore robust architectures and data augmentation strategies that maintain motion plausibility and stylistic coherence under noisy or incomplete input, thereby improving the reliability and generalization of music-to-dance generation systems.

\bibliographystyle{ACM-Reference-Format}
\bibliography{mm-paper}